\documentclass[prl,twocolumn,showpacs,superscriptaddress,preprintnumbers,amsmath,amssymb,amsfonts]{revtex4}

\usepackage{dcolumn}
\usepackage{bm}
\usepackage{times}
\usepackage[dvips]{graphicx}

\begin{document}

\preprint{APS/123-QED}

\title{Synchronization of coupled chaotic oscillators as a phase transition}
\author{F. T. Arecchi}
\email{arecchi@ino.it}
\affiliation{Istituto Nazionale di Ottica Applicata, Largo E. Fermi 6, 50125, Florence, Italy}
\affiliation{Department of Physics, University of  Florence, Italy}
\author{M. Ciszak}
\email{marzena@imedea.uib.es}
\affiliation{Istituto Nazionale di Ottica Applicata, Largo E. Fermi 6, 50125, Florence, Italy}
\affiliation{Department of Physics, University of  Balearic Islands, Spain}

\date{\today}

\begin{abstract}
We characterize the synchronization of an array of coupled chaotic elements as a phase transition where order parameters related to the joint probability at two sites obey power laws versus the mutual coupling strength; the phase transition corresponds to a change in the exponent of the power law. Since these studies are motivated by the behaviour of the cortical neurons in cognitive tasks, we account for the short time available to any brain decision by studying how the mutual coupling affects the transient behaviour of a synchronization transition over a fixed time interval. We present a novel feature, namely, the absence of decay of the initial defect density for small coupling.

\end{abstract}

\pacs{Valid PACS appear here}
\maketitle


Temporal coding vs. rate coding for the neural-based information has been open to debate in the neuroscience literature \cite{1}. In rate coding  only the   mean frequency of spikes  over a time interval matters, thus requiring a suitable counting interval, which seems unfit for  fast decision tasks. Temporal coding assigns importance to the precise timing and coordination of action potentials or ``spikes''. A special type of temporal coding is synchrony, whereby information is encoded by the synchronous firing of action potentials of all neurons of a cortical module. 

Let a laboratory animal be exposed to a visual field containing two separate objects, both made of the same visual elements. Since each receptive field isolates a specific detail of an object, we should expect a corresponding large set of individual responses. On the contrary, all the cortical neurons whose receptive fields are pointing to a specific object synchronize their spikes, and as a consequence the visual cortex organizes into separate neuron groups oscillating on two distinct spike trains for the two objects ({\em feature binding}). Experimental evidence was obtained by insertion in the cortical tissue of animals of microelectrodes each sensing a single neuron \cite{2}. Indirect evidence of synchronization has been reached for human subjects as well, by processing the EEG (electro-encephalo-gram) data \cite{3,4}.

The dynamics ruling the above facts should be described by a convenient dynamical model, independently of the biological components. Since feature binding   results from the readjustment of the temporal positions of the spikes, the most convenient strategy for this is to rely on the properties of suitable chaotic oscillators (c-o) \cite{5}.

In this paper we  explore the conditions under which an assembly of coupled c-o reaches mutual synchronization, thus displaying a coherent behaviour. We compare two dynamical models, namely HC \cite{5} and Roessler \cite{6}.  
Homoclinic chaos (HC) \cite{5} appears as the optimal strategy for a time code shared by a large crowd of identical coupled objects. Indeed HC provides at each pseudo-period (or inter spike interval = $ISI$) the alternation of a regular large spike (a) and a small chaotic background (b). (b) is the sensitive region where the activation from neighbors occurs, while the spike (a) provides a suitable signal to activate the coupling. Whence, a chain of weakly coupled c-o of this kind will easily synchronize, reaching a state common to all sites. 

We consider an array of identical c-o with nearest neighbour bidirectional coupling, ruled by \cite{10}
\begin{eqnarray}
\dot{x}_1^i&=&k_0x_1^i(x_2^i-1-k_1\sin ^2x_6^i) \nonumber\\
\dot{x}_2^i&=&-\gamma _1 x_2^i-2k_0x_1^ix_2^i+gx_3^i+x_4^i+p \nonumber \\
\dot{x}_3^i&=&-\gamma _1x_3^i+gx_2^i+x_5^i+p\nonumber \\
\dot{x}_4^i&=&-\gamma _2x_4^i+zx_2^i+gx_5^i+zp \\
\dot{x}_5^i&=&-\gamma _2x_5^i+zx_3^i+gx_4^i+zp\nonumber \\
\dot{x}_6^i&=&-\beta \left[x_6^i-b_0+r\left(f(x_1^i)+\epsilon(x_1^{i-1}+x_1^{i+1}-2\langle x_1^i\rangle)\right)\right] \nonumber 
\label{eq1} 
\end{eqnarray}
where $f(x_1^i)=\frac{x_1^i}{1+\alpha x_1^i}$. The index $i$ denotes the $i$th site position ($i=1,\hdots,N$), and dots denote temporal derivatives. The model \cite{10} was first introduced to describe the chaotic dynamics of a molecular ($CO_2$) laser with feedback. For each site $i$, variable $x_1$ represents the laser intensity, $x_2$ the population inversion between the two levels resonant with the radiation field, and $x_6$ the feedback voltage which controls the cavity losses. The auxiliary variables $x_3$, $x_4$ and $x_5$ account for molecular exchanges within the $CO_2$ molecule. We consider identical systems at each site $i$; as for the parameters, their physical meaning has been discussed elsewhere. Their values are:
$k_0=28.5714$, $k_1=4.5556$, $\gamma _1=10.0643$, $\gamma _2=1.0643$, $g=0.05$, $p_0=0.016$, $z=10$, $\beta =0.4286$, $\alpha =32.8767$, $r=160$ and $b_0=0.1032$. The mutual coupling consists of adding to the equation for $\dot{x}_6$ on each site a function of the intensity $x_1$ of the neighboring oscillators. The term $\langle x_1^i\rangle $ represents the average value of the $x_1^i$ variable, calculated as a moving average over a long time. The system is integrated by means of  Bulirsch-Stoer predictor-corrector method with open boundary conditions.

Any c-o where chaos is due to the homoclinic (or heteroclinic) return to a saddle focus displays a high sensitivity to an external perturbation just in the neighborhood of the saddle \cite{5}. We use the generic attribution of HC for systems which have a saddle focus instability \cite{11}. In view of this  high sensitivity, we expect that they synchronize not only under an external driving \cite{12}, but also for a convenient mutual coupling strength. A quantitative indicator of this sensitivity is represented by the so-called {\em propensity to synchronization} \cite{13} whereby different chaotic systems can be compared.
Furthermore, for an array of coupled c-o, the onset of a collective synchronization was explored numerically for some selected values of the coupling strength $\epsilon $, showing that the lack of synchronization manifests itself as phase slips which appear as dislocations in a space-time plot (space being the axis of site positions, time being the point-like occurrence of a spike) \cite{14}. Space–wise, a tiny variation of $\epsilon $ changes dramatically the size of the synchronized domain \cite{13}. Here we study in a systematic way the onset of synchronization. Introducing a suitable order paramater, we prove that it has a power law dependence on $\epsilon $ (control parameter); thus characterizing synchronization as a phase transition. 

\begin{figure}[h]
\includegraphics*[width=0.65\columnwidth]{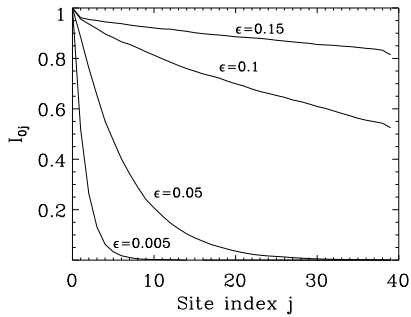}
\vspace{-0.5cm}
\caption{\label{fig1}Mutual information of two sites $i$ and $j$ versus the site separation $x=|i-j|$ for various amounts $\epsilon $ of the coupling strength.}
\end{figure}

\begin{figure}[h]
\hspace{0.25cm}(a)\hspace{4.cm}(b)\\
\includegraphics*[width=0.495\columnwidth]{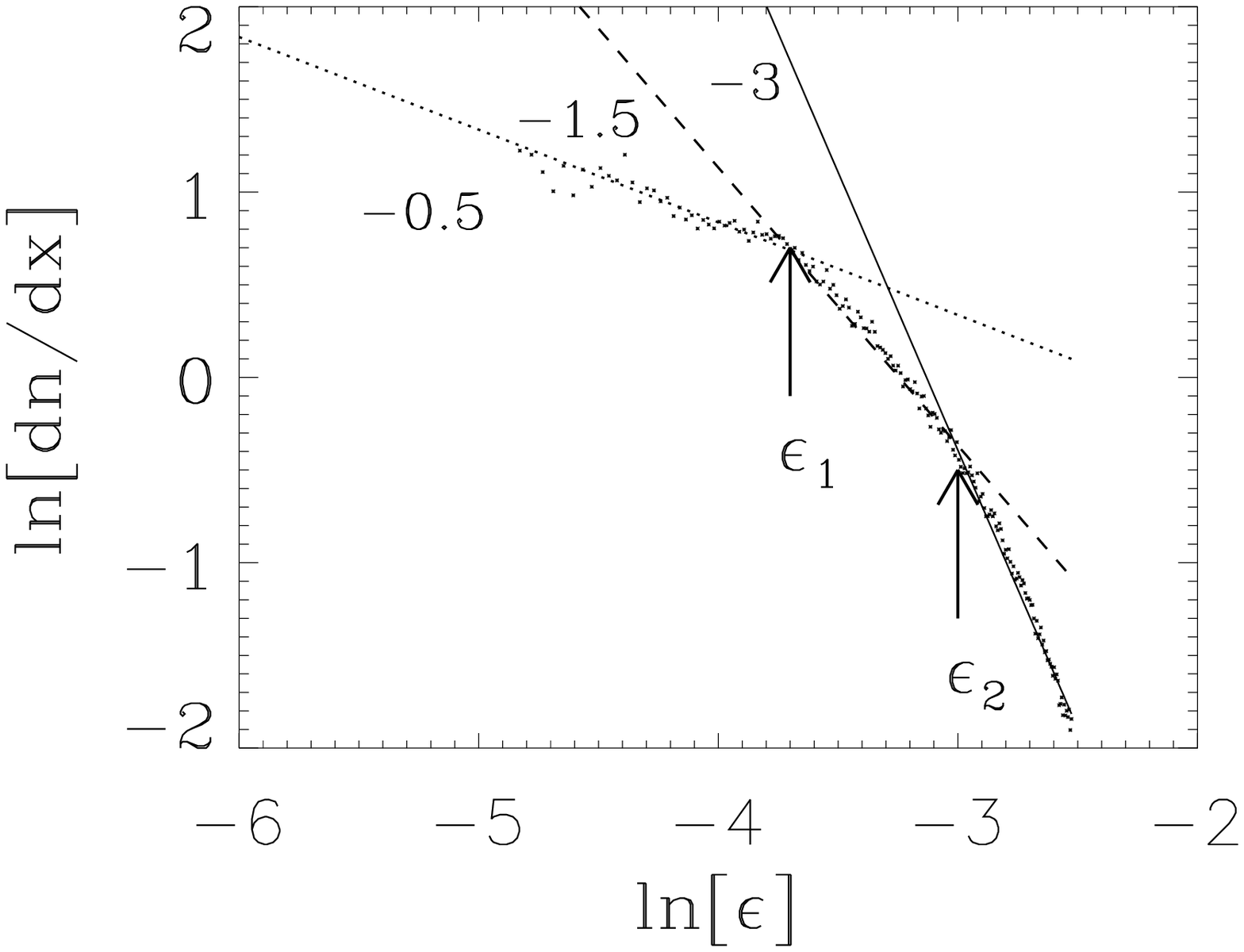}
\includegraphics*[width=0.495\columnwidth]{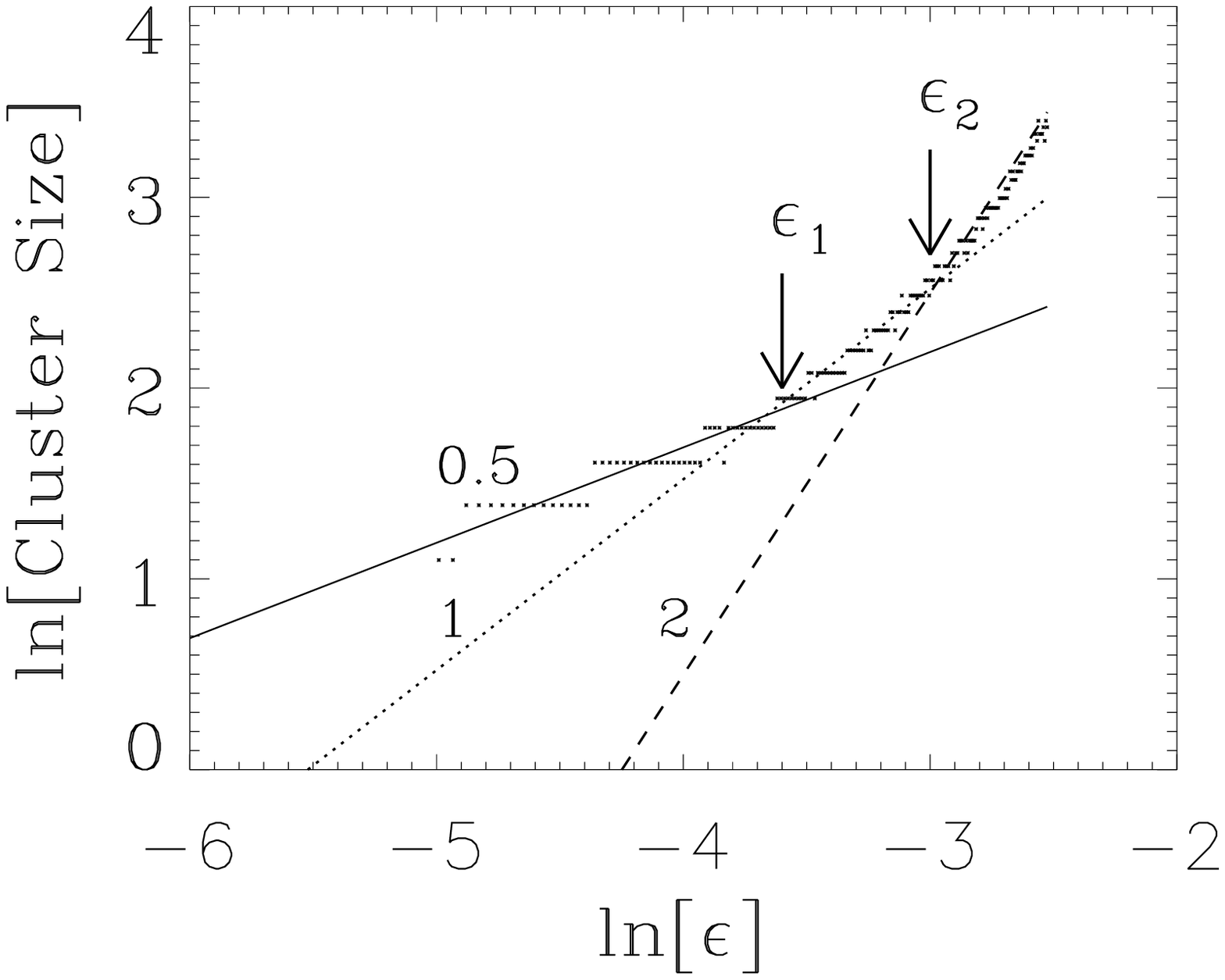}\\
\hspace{0.25cm}(c)\hspace{4.cm}(d)\\
\includegraphics*[width=0.495\columnwidth]{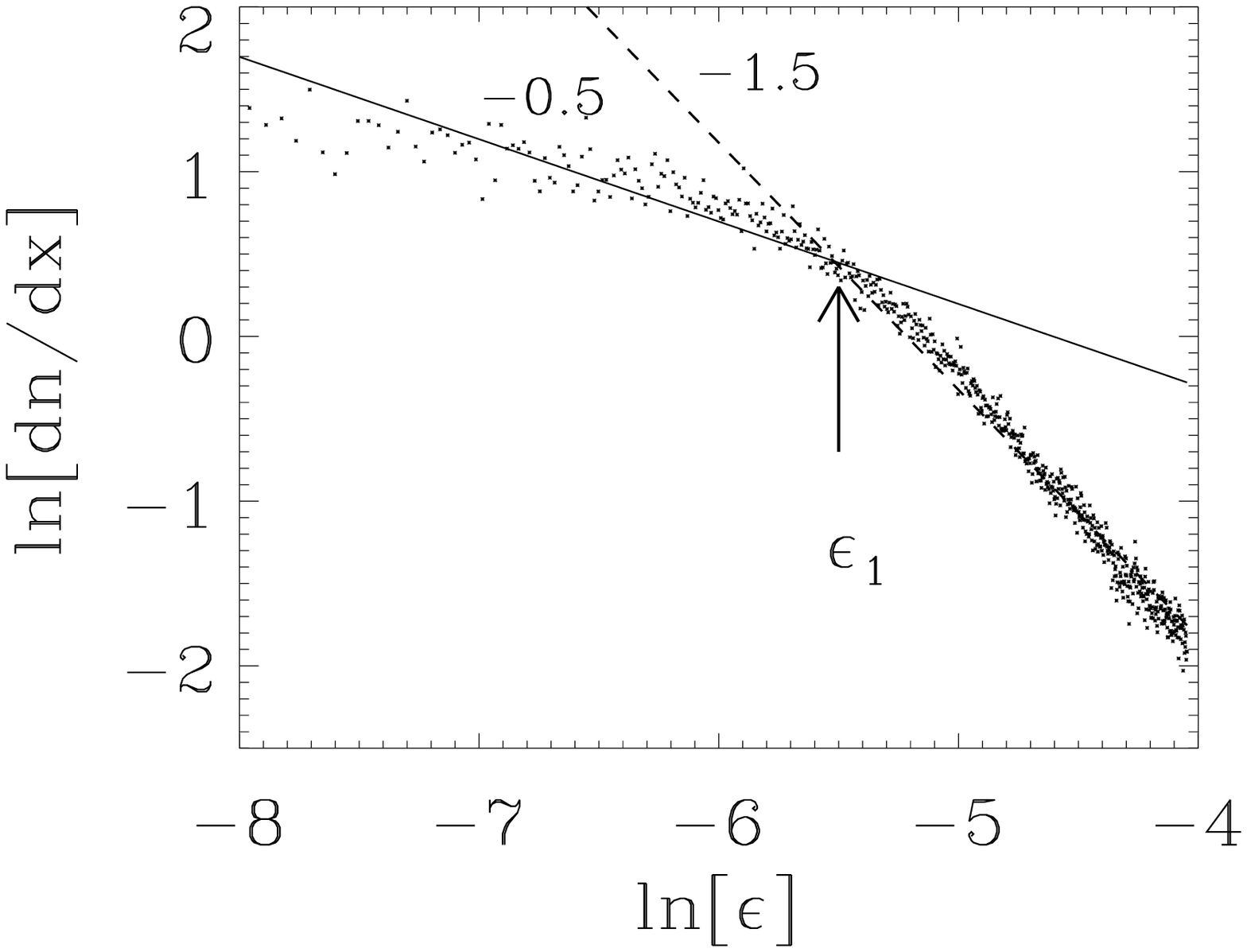}
\includegraphics*[width=0.495\columnwidth]{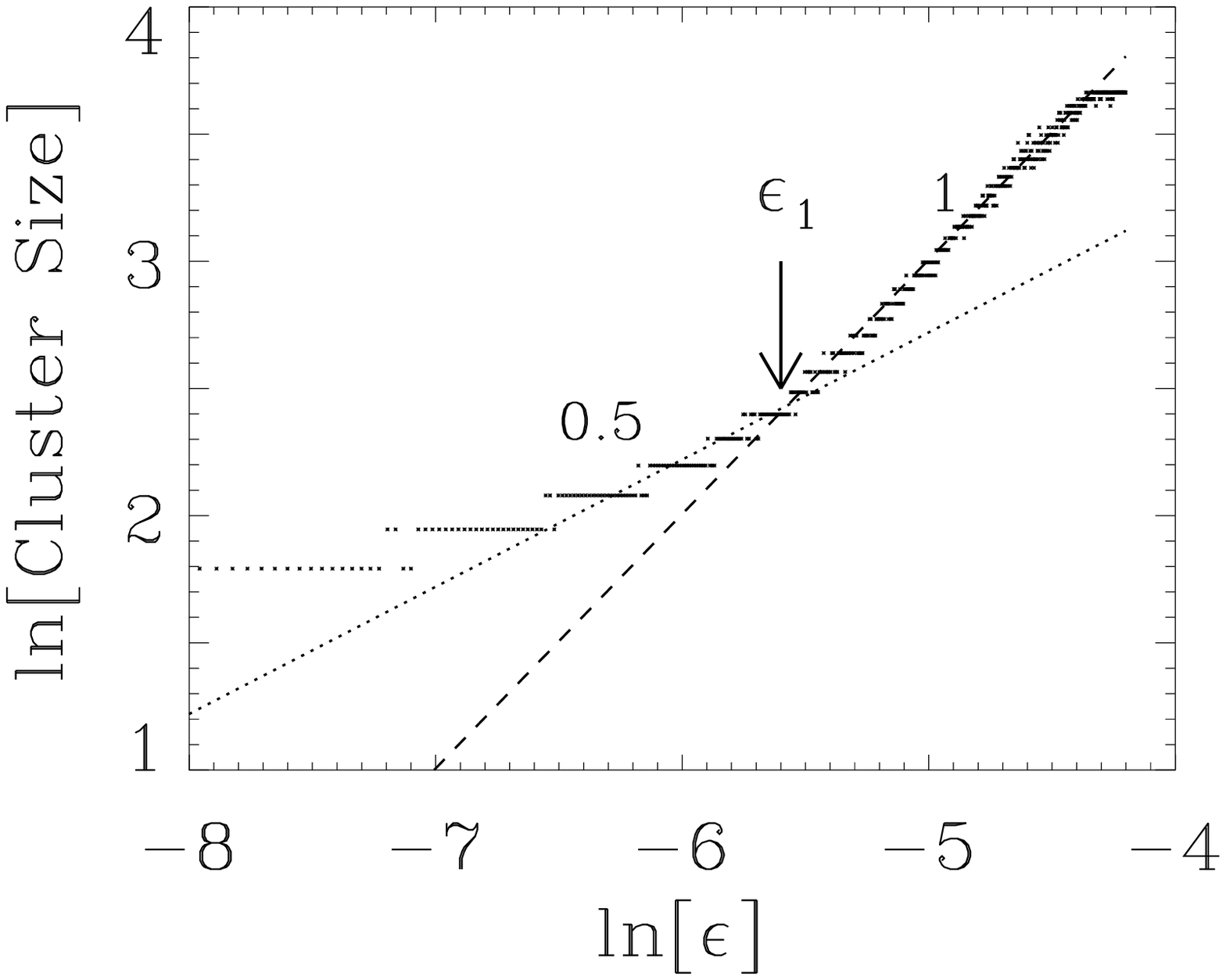}
\vspace{-0.3cm}
\caption{\label{fig2}Linear array of $40$ coupled HC c-o. (a) Average gradient of the cluster size $N$. It coincides with the gradient of the mutual information  (see Fig.~\ref{fig1}) averaged over all sites. The two cusp points are $\epsilon _1=0.0245$ and $\epsilon _2=0.049$. (b) As in a). Average cluster size versus $\epsilon $. The two cusp points are $\epsilon _1=0.0245$ and $\epsilon _2=0.049$ (the second one corresponding to saturation).(c) Array of $40$ Roessler c-o. The cluster size gradient displays a single cusp at $\epsilon _1=0.00405$. (d) As in c). Cluster size versus  $\epsilon $. There is only one cusp corresponding to saturation at $\epsilon _1=0.00405$. At variance with HC there is no intermediate speed up.}
\end{figure}

The most prominent feature of HC is that the time signal consists of almost identical spikes occuring at chaotic times, with an $\langle ISI\rangle $ (average interspike interval) much larger than the spike duration. Synchronization of two adjacent sites implies that pairs of spikes occuring on the two sites be separated by much less then $\langle ISI\rangle $. Precisely, when the spikes are separated by less then $10\% \langle ISI\rangle $ we will say that the two sites are locally synchronized and attribute a mutual probability of spike occurance $p_{i,i+1}=1$. If the separation is larger then  $10\% \langle ISI\rangle $ then we take $p_{i,i+1}=0$ and claim that locally synchronization is lost, because of a phase slip \cite{12}. 
Based on this set of joint probabilities we build a mutual information:
\begin{eqnarray}
I_{ij}=\sum p_{ij}\log \left(\frac{p_{ij}}{p_ip_j}\right)
\label{eq2}
\end{eqnarray}
Let us consider a linear array of $N$ nearest neighbours coupled sites. As we increase the coupling parameter $\epsilon $, the $I_{ij}$  vs the separation $|j-i|=x$ goes as Fig.~\ref{fig1}.
We plot the average gradient $\frac{dn}{dx}$ of $I_{ij}$ versus $\epsilon $ (Fig.~\ref{fig2}a). While initially $\frac{dn}{dx}$ goes as $\epsilon ^{-\frac{1}{2}}$, at $\epsilon = \epsilon _1$ there is a sharp transition point to a new power law dependence as $\epsilon ^{-\frac{3}{2}}$. Eventually at $\epsilon = \epsilon _2$ we have a second transition  to $\epsilon ^{-3}$. In order to give a meaning to the two transition points $\epsilon _{1,2}$, we plot the ``{\em cluster size}'' vs $\epsilon $. Using the method for cluster counting described in Ref. \cite{15}, two sites belong to the same cluster if $I_{ij}>\theta $  where $\theta $ is a fixed threshold. This way, we define ``{\em cluster size}'' as the size of the largest cluster which occurs within the array during the time evolution. The {\em cluster size} increases initially as $\epsilon ^\frac{1}{2}$, then as $\epsilon ^1$ between $\epsilon _1$ and $\epsilon _2$ and finally as $\epsilon ^2$ above $\epsilon = \epsilon _2$ (Fig.~\ref{fig2}b).
Now, an increase as $\epsilon ^\frac{1}{2}$ is typical of a linear diffusive process.  Indeed, the correlation domain in linear diffusion increases as $I_c \propto \sqrt{\epsilon t}$, thus, for a fixed observation time t, we have the power law $I_c \propto \epsilon ^\frac{1}{2}$. We therefore consider the steep increase beyond the transition point $\epsilon >\epsilon _2$ as a nontrivial nonlinear effect. We take the change of slope at $\epsilon _1$ as the onset of synchronization. 
We wish to show that this steep synchronization is a peculiarity of HC systems. For this purpose, we study an array of chaotic oscillators where no {\em spike}, but rather {\em phase synchronization} occurs. Such is the case of coupled Roessler oscillators \cite{6}, where  the chaotic oscillations are characterized by a main oscillation frequency,  and chaos appears as a spread in the amplitude of different cycles; here in fact the array tends toward a phase synchronization. The dynamics is ruled by \cite{16}
\begin{eqnarray}
\dot{x}_1^i&=&-x_2^i-x_3^i+\epsilon (x_1^{i-1}+x_1^{i+1}-2x_1^i)\nonumber \\
\dot{x}_2^i&=&x_1^i+ax_2^i \nonumber \\
\dot{x}_3^i&=&b+x_3^i(x_1^i-c)\label{eq3}
\end{eqnarray}
where $a=0.15$, $b=0.2$ and $c=10$. Independently from the chaotic amplitude, the phase difference at two different sites can vary smoothly from $0$ to $\pi$. Calling $T=\frac{2\pi }{\omega _0}$ the period of the main frequency $\omega _0$, then the joint probability of two sites can be taken as:
\begin{eqnarray}
p_{ij}=\cos \pi \frac{T_{ij}}{T}\label{eq4}
\end{eqnarray}
and it goes with continuity from $1$ ($T_{ij}=0$) to $0$ ($T_{ij}=\pm \frac{T}{2}$). Using such a probability we evaluate the mutual information and the cluster size, Fig.~\ref{fig2}c and Fig.~\ref{fig2}d. 
\begin{figure}[h]
\hspace{0.2cm}(a)\hspace{3.5cm}(b)\\
\includegraphics*[width=0.265\columnwidth]{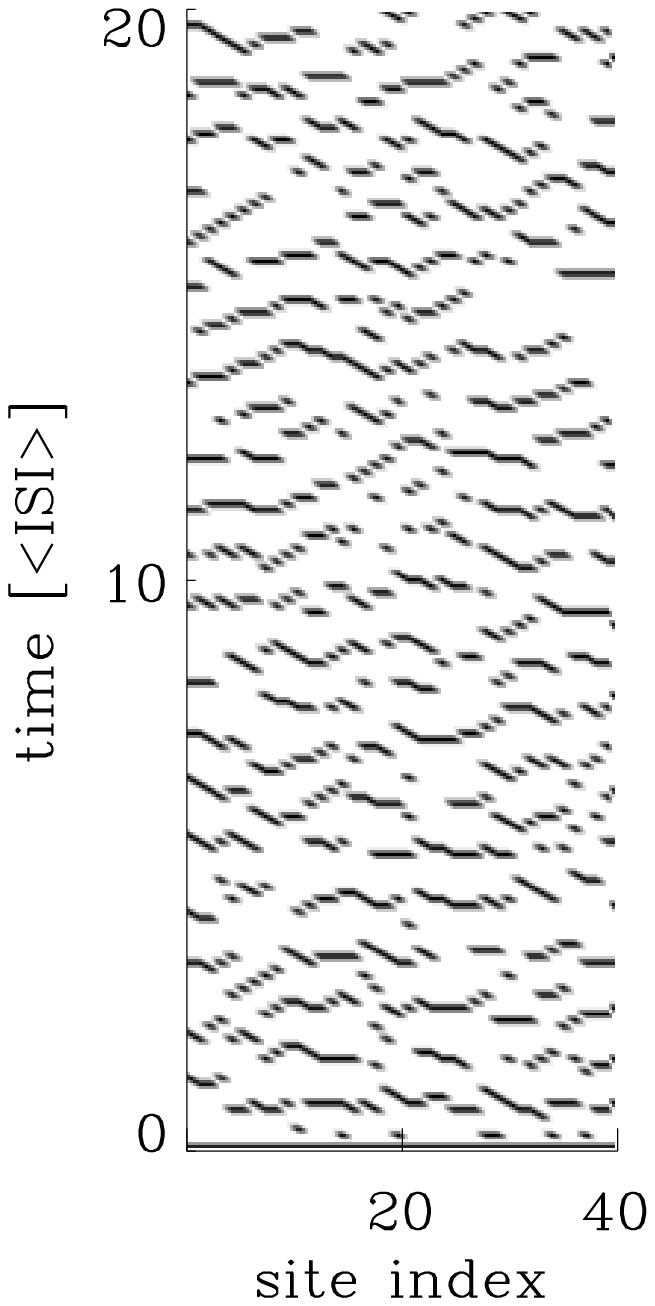}
\includegraphics*[width=0.235\columnwidth]{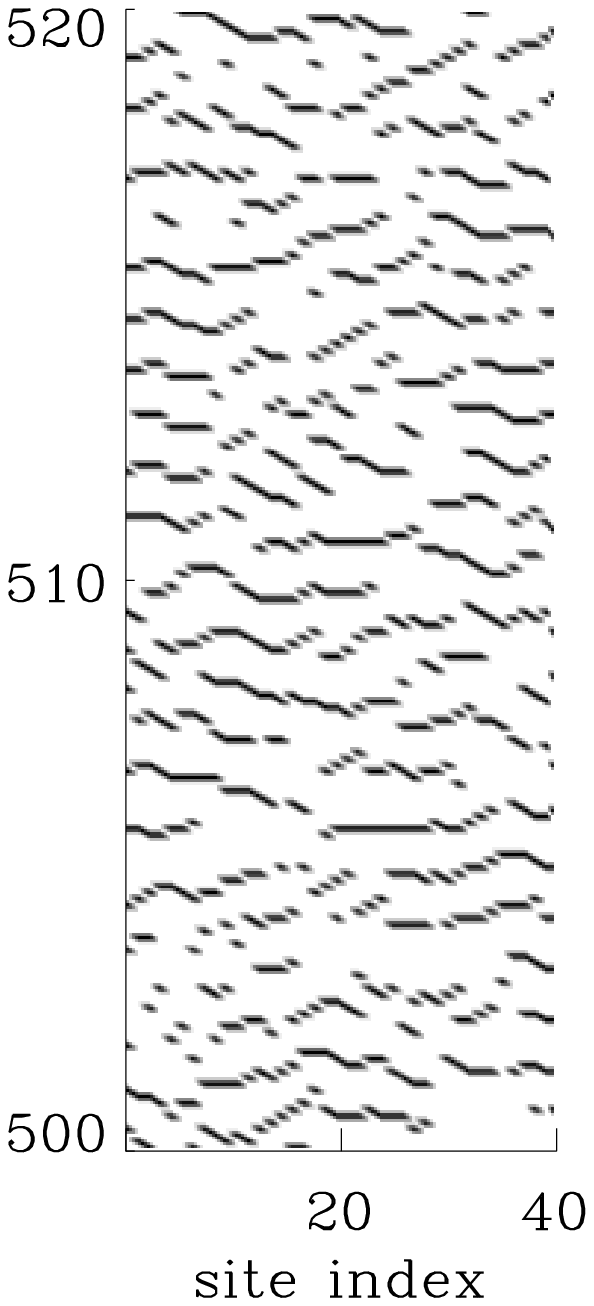}
\includegraphics*[width=0.215\columnwidth]{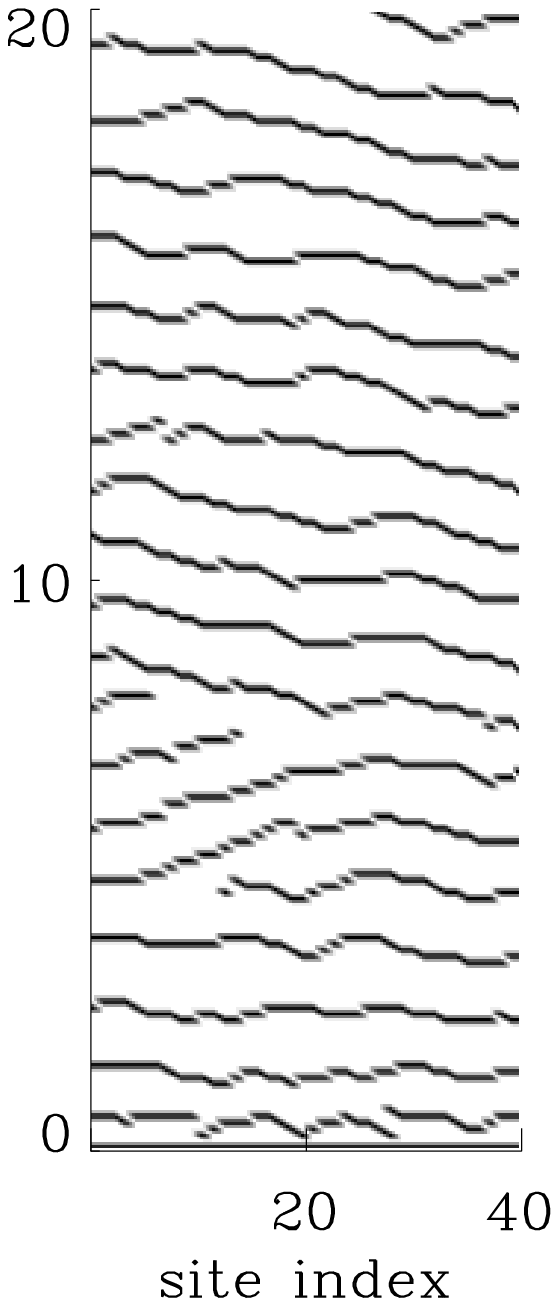}
\includegraphics*[width=0.235\columnwidth]{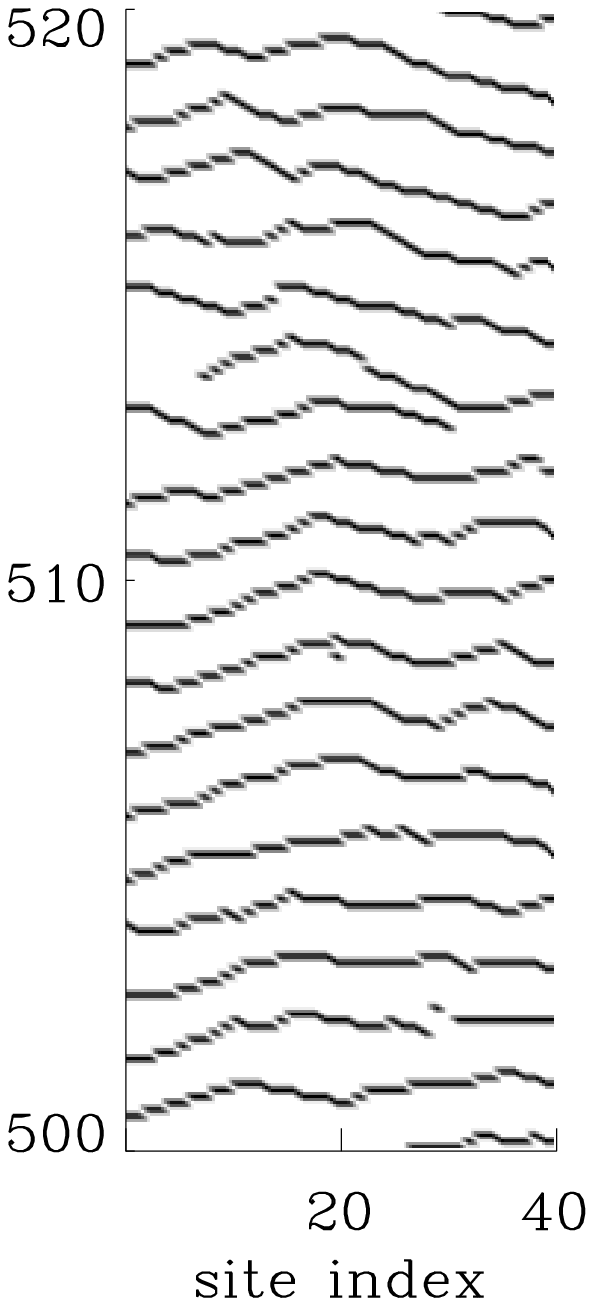}
\vspace{-0.3cm}
\caption{\label{fig3} Spatiotemporal evolutions of  HC systems for coupling constant (a) $\epsilon =0.03 $ and (b) $\epsilon =0.1$; time in the vertical axis is given in $\langle ISI\rangle $ units.}
\end{figure}
They show that $\epsilon _2$ is missing, and above $\epsilon _1$ the average gradient changes from $\epsilon ^{-\frac{1}{2}}$ to $\epsilon ^{-\frac{3}{2}}$. Notice that, up to total invasion of the available domain, the progress of phase synchronization grows as $\epsilon ^{\frac{1}{2}}$,  as in linear diffusion.

This fact confirms that the steep rise of synchronization beyond $\epsilon _2$ is a peculiar aspect of HC, or, more generally, of any chaotic dynamical system where chaos is due to the homoclinic return to a saddle focus. We use the generic attribution of HC for a large class of systems, with a saddle focus instability \cite{11}. This includes a class B laser with feedback \cite{10}, the Hodgkin-Huxley model for the production of action potentials in a neuron membrane \cite{17} and the Hindmarsh-Rose model of generation of spike bursts \cite{18}.
As a partial conclusion,  HC is a very plausible model to explain {\em feature binding} in an array of neurons belonging to the same cortical module.
\begin{figure}[h]
\hspace{0.25cm}(a)\hspace{4.cm}(b)\\
\includegraphics*[width=0.495\columnwidth]{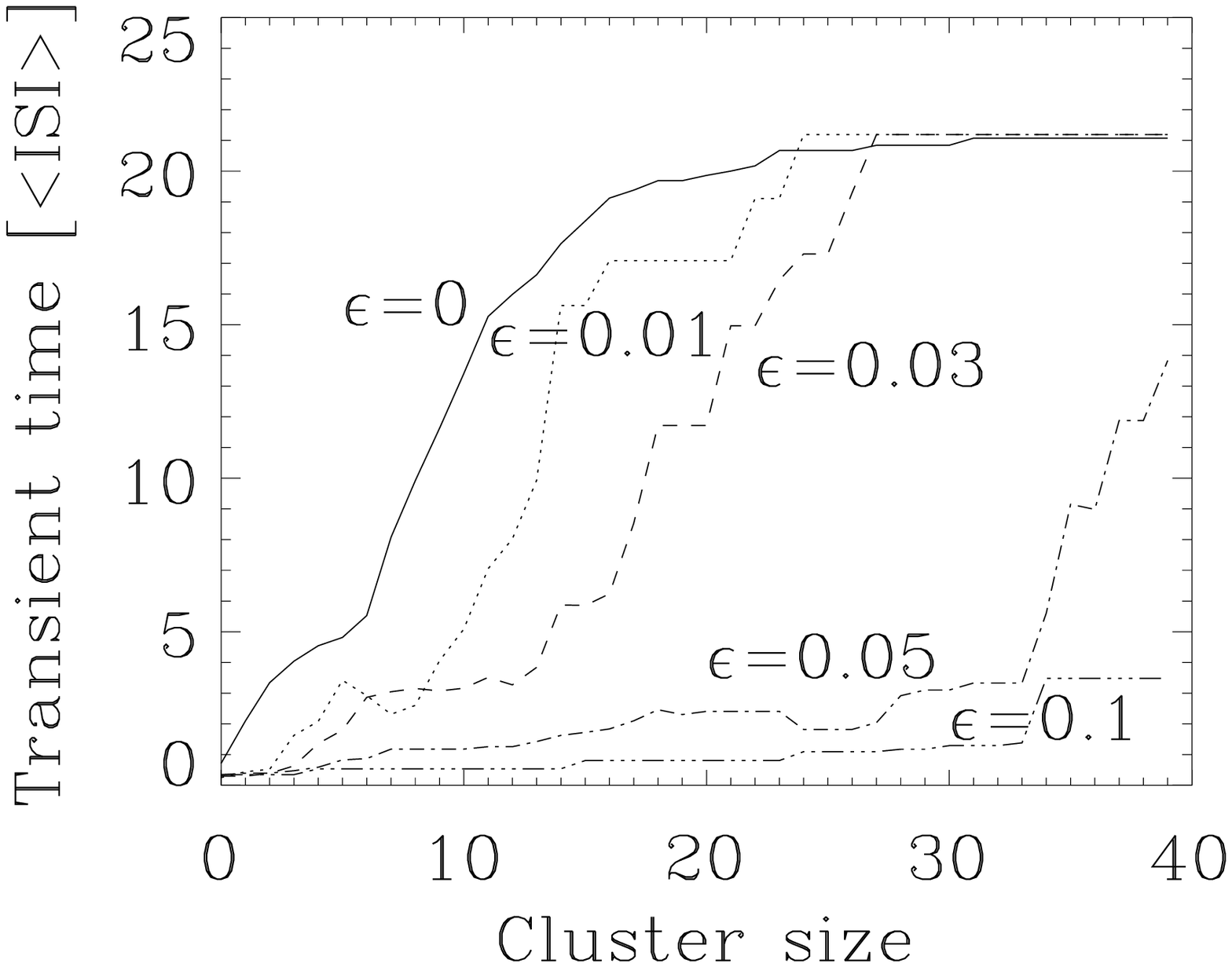}
\includegraphics*[width=0.495\columnwidth]{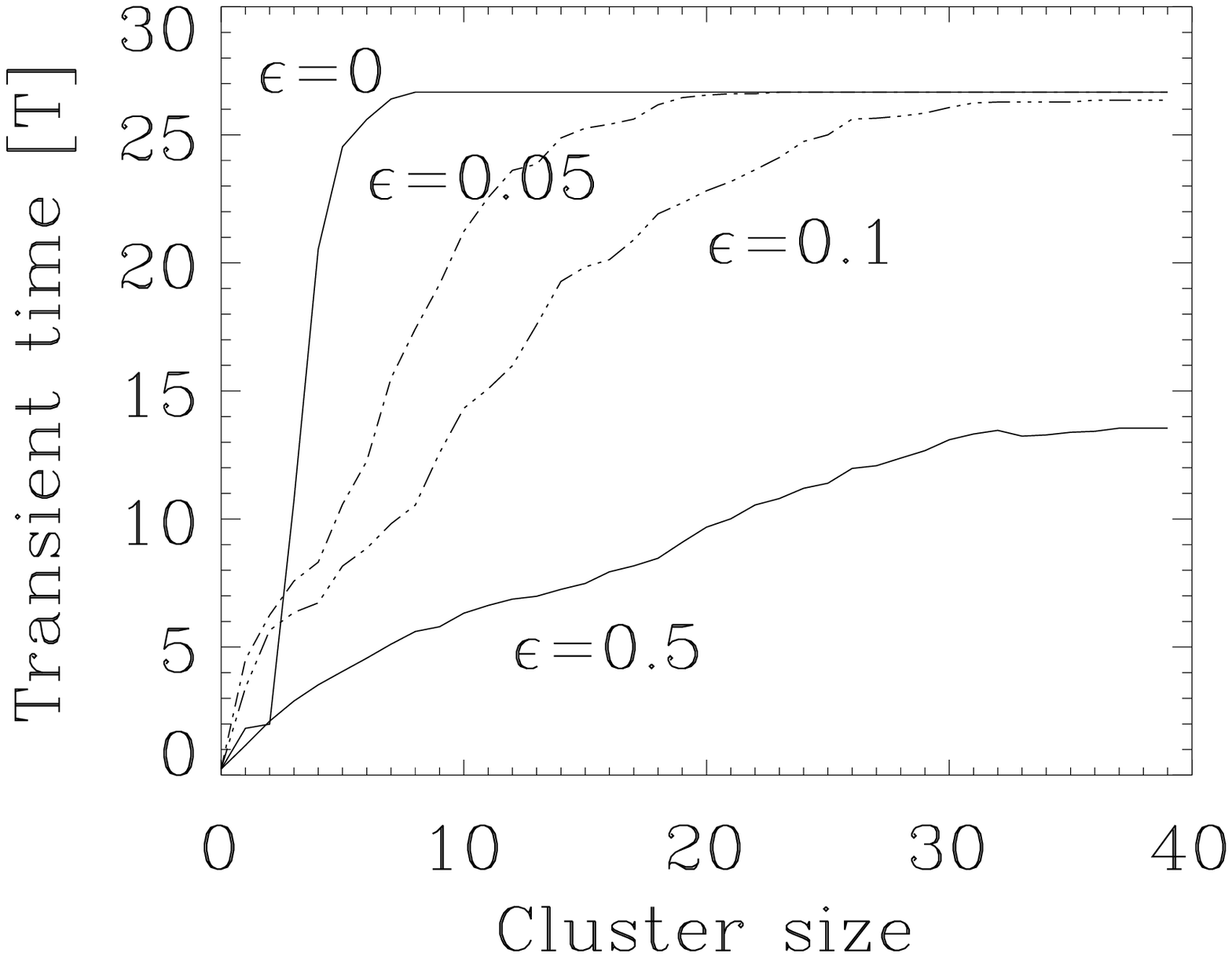}
\vspace{-0.3cm}
\caption{\label{fig4} Transient times versus cluster size for (a) HC and (b) Roessler c-o calculated for selected values of coupling constant during the limited time period taken as approximately $20$ spikes. Transient time is represented in the units of the average interspike interval  $\langle ISI\rangle $ in the case of HC and in units of constant period $T$ in the case of Roessler.}
\end{figure}
\begin{figure}[h]
\hspace{0.2cm}(a)\\
\hspace{-0.4cm}
\includegraphics*[width=0.7\columnwidth]{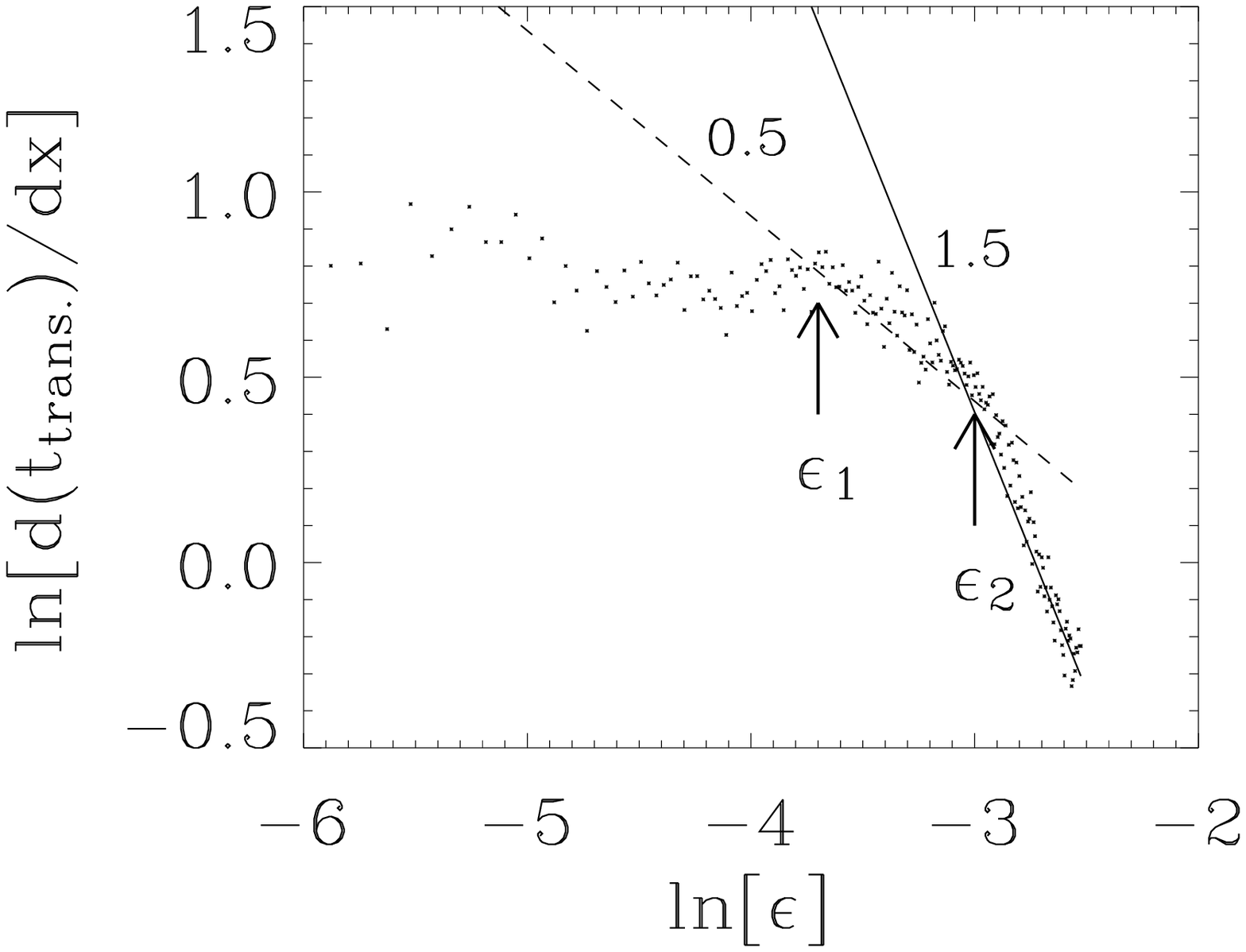}\\
(b)\\
\includegraphics*[width=0.7\columnwidth]{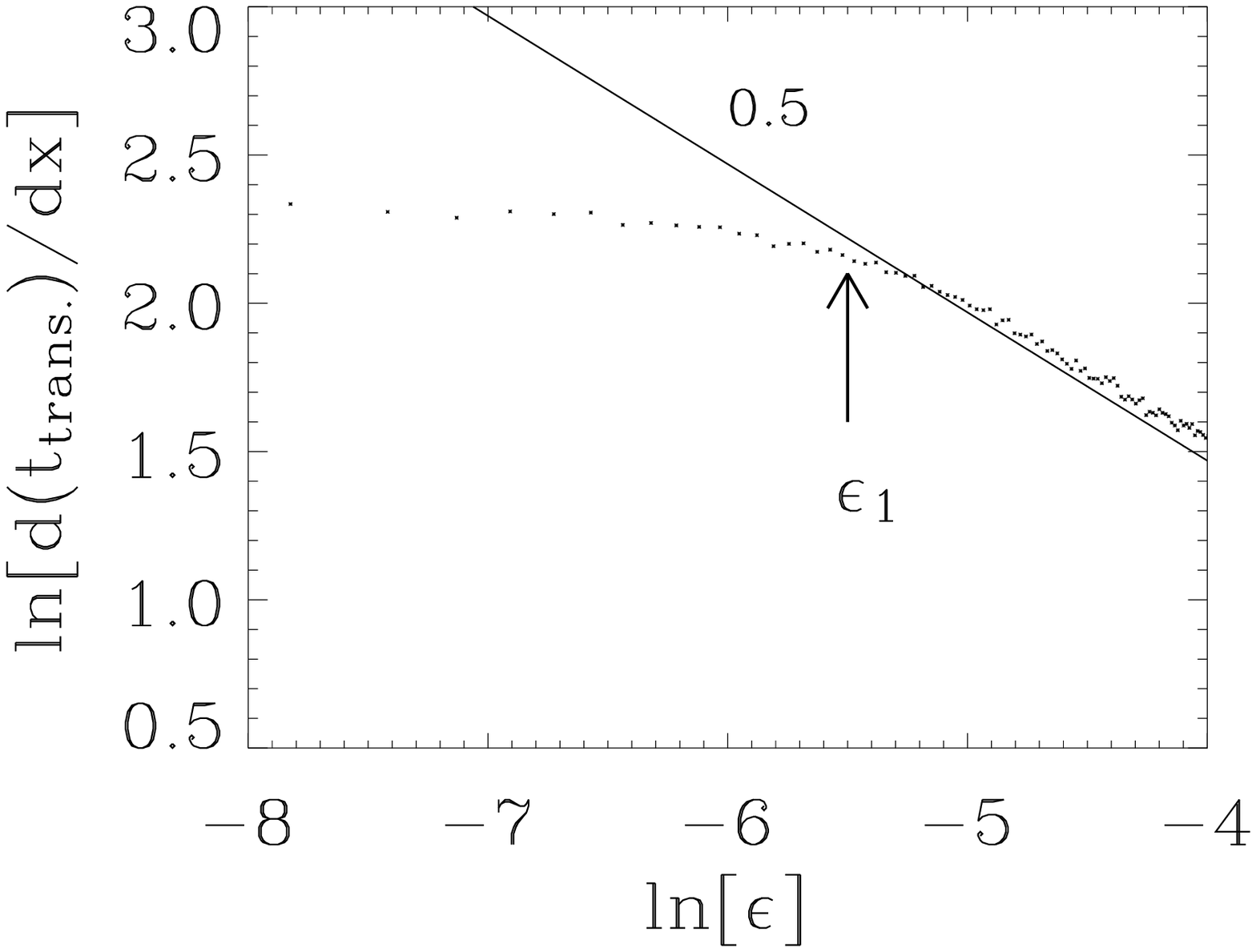}
\vspace{-0.3cm}
\caption{\label{fig5} Linear array of $40$ coupled HC (a) and Roessler (b) c-o. Average gradient of the transient time $t_{trans}$ versus coupling $\epsilon $. The two cusp points in (a) are at $\epsilon _1=0.0245$ and $\epsilon _2=0.049$ and the cusp in (b) is at $\epsilon _1=0.00405$. The cusp points in both cases coincide with those of Fig.~\ref{fig2}.}
\end{figure}

The onset of synchronization, either spike in HC or phase in Roessler, is described by order parameters (the gradient of the mutual information  and the cluster size) which undergo a phase transition as functions of a control parameter (the mutual coupling strength). The signature of the phase transition is the change of the power law behaviour, evaluated for long time well beyond any initial transient.

However, a perceptual task must be completed within a short time after the arrival of external stimuli, in order for the living organism to take  vital decisions as a response to those stimuli. This time for human subjects is around $200$ $ms$  which corresponds to about $10\langle ISI\rangle $ in the so called $\gamma $ band which provides relevant cues in visual tasks \cite{6}.

We then consider a {\em transient phase transition}, requiring that all sites be synchronized within less than  $20\langle ISI\rangle $. Notice that the criterion given above for $p_{i,i+1}=1$ in HC {\em does not} identify synchronization with isochronism, but it just rules out a phase slip (one spike more or less) between site $i$ and $j$. In a space-time plot where spike occurrence is denoted by a dot, a phase slip would appear as a dislocation (lack of connectedness) in an array of lines joining the spike occurrences at different sites \cite{13}.
Space-time plots are shown in Fig.~\ref{fig3}a ($\epsilon =0.03$) and Fig.~\ref{fig3}b ($\epsilon =0.1$) where the time is given  in  $\langle ISI\rangle $ units; the plots show time intervals starting from $t=0$ as well as from $t=500$.
 
For $\epsilon =0.03$, the average number of defects does not decay even for long times; indeed a plot of the same interval starting from $t=500$ displays the same number of defects on average; defects die out at one site and are born again at another one in course of time, due to the temporal relations between adjacent sites necessary to induce the escape from the saddle region (see the detailed discussion for unidirectional coupling in Ref. \cite{19}). On the contrary, for $\epsilon =0.1$, defects   decays within  $20\langle ISI\rangle $.

In Fig.~\ref{fig4} we show the transient time (in $\langle ISI\rangle $ units) necessary to reach a cluster size $x$ ($x$ can vary from $1$ to $40$), calculated for an ensemble of random initial conditions. The apparent monotonic increase with time of the cluster size for low $\epsilon $ is an ensemble averaging contrivance, irrelevant for the single trial. Having the distribution of transient times over the cluster sizes for different couplings we consider the averaged space gradient $\frac{d(t_{trans})}{dx}$. This quantity yields information on how the transient times needed to reach a given cluster size changes with the coupling and provides a global picture of the transient dynamics in the spatiotemporal system. For low $\epsilon $ we have a flat dependence on $\epsilon $ which means that the coupling is not effective at all in inducing synchronization, neither in HC (Fig.~\ref{fig5}a) nor in Roessler (Fig.~\ref{fig5}b); the critical values  $\epsilon _1=0.0245$ and  $\epsilon _2=0.049$  sign the starting point for a power law decay of that gradient in HC; similarly does $\epsilon _1=0.00405$ for Roessler. 

Let us elaborate on the persistence of initial defects and show that it is peculiar of synchronized HC arrays. In HC a defect corresponds to a sudden jump of $p_{ij}$ from $1$ to $0$ or viceversa. In order to investigate this new feature in another system, we   introduce a symbolic dynamics in Roessler as well. In Eq. \ref{eq4} $p_{ij}$ goes smoothly from $1$ at $T_{ij}=0$ to $0$ at $T_{ij}=\frac{T}{2}$;  we replace that smooth behaviour with a jumpy one, by taking $p_{ij}=1$ when $T_{ij} >\frac{1}{2}$ and $p_{ij} =0$ when $T_{ij} <\frac{1}{2}$.
This way, we obtain space-time plots of defects for Roessler which resemble those of Fig.~\ref{fig3}. However, here defects always decay for long times, as checked  for $\epsilon =0.001$ and $\epsilon =0.0001$, and thus the array is fully synchronized after a finite time, even for very small $\epsilon $. The flat region below $\epsilon _1=0.00405$ in Fig.~\ref{fig5}b is due to the fact that we were studying the transition to synchronization for short times, up to $20T$. Thus, there exists a crucial difference between HC and Roessler: for small $\epsilon $ in HC, even after reaching the synchronization of the whole array, the defects may reappear and then again disappear, on the contrary in Roessler, once full synchronization is reached, the defects do not reappear. Transient defects considered in sudden symmetry changes of Ginzburg-Landau type systems decay always, as shown theoretically \cite{20} and verified experimentally in He3 \cite{21} and in nonlinear optics \cite{22}. Instead, below synchronization threshold of an HC array, there is a steady defect sea. It would be interesting to explore the appearance of such a feature in situations different from that here considered. Furthermore, this report has been limited to spontaneous synchronization; from a biological point of view it is crucial to explore the general scaling laws in presence of external forcing as already done, but only for some numerical examples \cite{14}.

This work was partially supported by Ente Cassa di Risparmio di Firenze under the Project “Dinamiche cerebrali caotiche” and by the EU under the ENOC COST action.

\end{document}